# Kochen–Specker non-contextuality through the lens of quantization

Simon Friederich *, Mritunjay Tyagi

*University of Groningen, University College Groningen, The Netherlands*

ABSTRACT

The Kochen–Specker theorem shows that it is impossible to assign sharp values to all dynamical variables in quantum mechanics in such a way that the algebraic relations among the values of dynamical variables whose self-adjoint operators commute are the same as those among the operators themselves. We point out that, for quantum theories obtained by *quantizing* some classical theory, this condition –*Kochen–Specker non-contextuality* – is implausible from the start because quantization usually changes algebraic relations. We explain why this is so, using the formalism of deformation quantization and its conception of *star products*, and we illustrate the relevance of this point using various examples of dynamical variables quantized via Weyl quantization and coherent state quantization. Our observations suggest that the relevance of the Kochen–Specker theorem to the question of whether one can assign sharp values to all dynamical variables is rather limited.

## 1. Introduction

The Kochen–Specker theorem (sometimes called Bell–Kochen–Specker theorem) [1,2], a celebrated result in the foundations of quantum theory, establishes that it is impossible to assign sharp values to all dynamical variables of a quantum system with Hilbert space dimension $\geq 3$ in such a way that the algebraic relations among the values assigned to dynamical variables whose self-adjoint operators commute are the same as those among the operators themselves. Since it seems to many that natural and uncontrived assignments of sharp values of dynamical variables should have this feature, the Kochen–Specker theorem is typically seen as speaking against the very idea of assigning sharp values to all dynamical variables. The theorem strengthens an older result by von Neumann [3], famously criticized by Hermann [4] and Bell [2], which did not restrict the condition that the algebraic relations among values and operators should match to commuting operators. The condition on hypothetical assignments of sharp values imposed by Kochen and Specker has been termed (Kochen–Specker) *non-contextuality*.

In this paper, we argue that, for quantum theories with self-adjoint operators obtained via *quantization* of some class of dynamical variables defined on phase space, Kochen–Specker non-contextuality is implausible from the start as an assumption about the values of these variables. As far as we are aware, our argument is new. The potential relevance of quantization for the plausibility of Kochen–Specker non-contextuality was once pointed out by Isham [5, p. 195], but Isham does not seem to have considered the implications of this insight, which will be spelt out in this paper.

A sketch of our argument, which, after a review of the Kochen–Specker theorem in Section 2, is to be presented in Sections 3–5 of this paper, is as follows:

Imagine a hypothetical system that has an exact position in phase space. Like a typical classical system, it has clearly defined dynamical variables, which are functions on phase space chosen from an appropriate set, for example the real-valued measurable functions. We may construct a quantum theory by mapping these functions to self-adjoint Hilbert space operators via some specific quantization method. In this perspective, the self-adjoint operators are not themselves the dynamical variables, but rather *represent* the genuine dynamical variables – the phase-space functions – for the purposes of computing expectation values or time evolution.

* Corresponding author.
*E-mail address:* s.m.friederich@rug.nl (S. Friederich).

There is an important reason for adopting this perspective. The quantities to which experimentalists have direct access are always phase space-type quantities: the positions of pointer needles, the coordinates of spots on screens, or, more generally, the macroscopic configuration of the measurement device. These are naturally described in classical terms, as positions (and perhaps momenta) of macroscopic objects, not as values of abstract self-adjoint operators. In the scenario which we explore – one in which all quantum systems possess sharp phase-space locations – it is natural to extend this account to the microscopic domain as well. On this view, the dynamical variables of microscopic systems are simply their phase-space functions, and the associated operators merely represent these functions in the Hilbert-space formalism for the sake of computational convenience.

Alternatively, if one prefers to treat the self-adjoint operators as the genuine "quantum" dynamical variables, the same considerations suggest to use the reverse of the quantization mapping, sometimes called *de-quantization*, to the extent that it is unique, as the proper procedure to assign values to the self-adjoint operators.[1] The phase space function mapped to a given self-adjoint operator is called the "symbol" of the operator with respect to the chosen quantization scheme. Put differently, we propose that the value assigned to a self-adjoint operator be taken as its *symbol* under some suitably chosen quantization scheme, evaluated at a specific phase space location.

Next, we observe that, in the leading approaches to quantization, quantization does not commute with functional composition.[2] Notably, quantization does not leave algebraic relations invariant (nor does de-quantization). Concretely, as we show, for the two (arguably) most popular quantization schemes, namely, Weyl quantization and coherent state ("anti-Wick") quantization, algebraic relations among dynamical variables do *not* in general match algebraic relations among self-adjoint operators. Notably, the algebraic relations between projection operators, which figure crucially in the proof of the Kochen–Specker theorem, do not in general match those among the phase space functions assigned to them via quantization. Accordingly, a class of very natural assignments of sharp values to dynamical variables – namely, those, that assign sharp values via (de-) quantization – violates Kochen–Specker non-contextuality from the very start. This challenges the received wisdom according to which the Kochen–Specker theorem indicates that only rather contrived and unnatural assignments of sharp values to dynamical variables are possible.

We conclude the paper in Section 6 by considering in a little more detail what it means to combine the assignment of sharp values to all dynamical values with coherent state quantization specifically. For coherent state quantization it turns out to be possible to interpret the expectation values of dynamical variables $A$ computed quantum mechanically as $Tr(\hat{\rho}\hat{A})$, with $\hat{A}$ obtained from its "symbol" $A$ via coherent state quantization, as phase space averages $\int A(\mathbf{x},\mathbf{p})\, F(\mathbf{x},\mathbf{p})\, d\mathbf{x}\, d\mathbf{p}$. The role of the phase space probability density $F(\mathbf{x},\mathbf{p})$ is played by the Husimi function [7]. The idea to regard the Husimi function as a proper probability density (and not merely a "quasi-probability function") seems to have been first proposed by Fritz Bopp [8] and has recently been suggested and advocated again by Drummond and Reid [9] as well as Friederich [10].

It should be noted that our criticism of Kochen–Specker non-contextuality does not apply to notions of non-contextuality beyond Kochen–Specker non-contextuality which have also been suggested in the literature, see [11] for an overview and categorization. Unlike Kochen–Specker non-contextuality, these tend to be dynamical notions, often phrased in terms of operational vocabulary (see [12] for criticism), treating preparations, processes and measurements separately [13]. A limitation of these notions is that they presuppose the existence of preparation-independent response functions, i.e. they assume that the ontic state $\lambda$ of the system prior to measurement screens off any correlation between the preparation and the measurement ("$\lambda$-mediation" [14, Assumption V.5], "$\lambda$-sufficiency" [11]). Adlam [15], in our view persuasively, cautions against taking this assumption for granted when developing single world-realist accounts of quantum theory. Kochen–Specker non-contextuality, in contrast to the dynamical accounts of non-contextuality, does not rely on any assumptions about the dynamics of the values of dynamical variables and is in that sense more widely applicable.

## 2. Recapitulating the Kochen–Specker theorem

Kochen and Specker's own statement and proof of their theorem as given in [1] is based on the notion of a *partial algebra*. The technicalities related to that notion are largely avoided in the more pedagogical account given by Held [16], which suffices for the present discussion. The theorem itself, as stated there, is as follows (notation adjusted in what follows, to make it clear that "dynamical variables", in Held's writing, refers to self-adjoint Hilbert space operators, whereas elsewhere in this paper we use "dynamical variable" to refer to phase space functions which, in turn, are "represented" by the self-adjoint operators on which they are mapped via quantization):

> Let $H$ be a Hilbert space of QM state vectors of dimension $x \geq 3$. There is a set $M$ of observables on $H$, containing $y$ elements, such that the following two assumptions are contradictory:
>
> (KS1) All $y$ members of $M$ simultaneously have values, i.e. are unambiguously mapped onto real numbers (designated, for observables $\hat{A}$, $\hat{B}$, $\hat{C}$, $\ldots$, by $v(\hat{A})$, $v(\hat{B})$, $v(\hat{C})$, $\ldots$).

[1] De-quantization must be sharply distinguished from "taking the classical limit". De-quantization is simply the inverse (if it exists) of the quantization mapping from phase space functions to self-adjoint operators. Taking the classical limit means considering the quantum theory in a regime (if one exists) where a classical description provides a good approximation. De-quantization may, but need not, be part of considering the classical limit.

[2] See the discussion of "condition (q3)" on quantization schemes and its generic failure in [6, Sect. 1].

(KS2) Values of all observables in $M$ conform to the following constraints:

(a) If $\hat{A}$, $\hat{B}$, $\hat{C}$ are all compatible [i.e. pairwise commute] and $\hat{C} = \hat{A} + \hat{B}$, then $v(\hat{C}) = v(\hat{A}) + v(\hat{B})$;

(b) if $\hat{A}$, $\hat{B}$, $\hat{C}$ are all compatible and $\hat{C} = \hat{A}\hat{B}$, then $v(\hat{C}) = v(\hat{A}) \cdot v(\hat{B})$. Held [16, Sect. 3.1]

The proof of the Kochen–Specker theorem focuses specifically on projection operators. (In fact, Bell's proof of a similar result [2] does not appeal to any operators beyond projection operators at all, see [17] for a useful clarification of the relation between the proofs and what they establish.)

Suppose that $\{|x_j\rangle\}$ is an orthonormal basis of $H$ with $j = 1, \ldots, n$, where the Hilbert space dimension $n$ is at least 3 and possibly infinite. The projection operators $|x_j\rangle\langle x_j|$ resolve the identity: $\sum_{j=1}^{n} |x_j\rangle\langle x_j| = \mathbf{1}$. So, by KS2 (a),

$$\sum_{j=1}^{n} v(|x_j\rangle\langle x_j|) = v(\mathbf{1}) \,, \tag{1}$$

According to KS2 (b), $v(\mathbf{1})$ must be 1 because, if $\hat{R}$ is some arbitrary self-adjoint operator, then $v(\hat{R}) = v(\hat{R}\mathbf{1}) = v(\hat{R}) \cdot v(\mathbf{1})$, which implies $v(\mathbf{1}) = 1$.

Moreover, KS2 (b) implies that, for every $j$, since $|x_j\rangle\langle x_j|$ is a projection operator, which means $(|x_j\rangle\langle x_j|)^2 = |x_j\rangle\langle x_j|$,

$$\left(v(|x_j\rangle\langle x_j|)\right)^2 = v\left((|x_j\rangle\langle x_j|)^2\right) = v(|x_j\rangle\langle x_j|) \,. \tag{2}$$

Accordingly, any $v(|x_j\rangle\langle x_j|)$ is either 0 or 1.

We can conclude that, for any orthonormal basis $\{|x_j\rangle\}$, the values $v(|x_j\rangle\langle x_j|)$ are all zero, except for one index $j$, for which $v(|x_j\rangle\langle x_j|) = 1$.

From this point, the proof of the Kochen–Specker theorem is an exercise in assigning numbers 0 and 1 to projection operators in such a way that the requirement just established holds for arbitrary orthonormal bases $\{|x_j\rangle\}$ and showing that this is in general impossible if the dimension of $H$ is $\geq 3$. We will not focus on that proof stage in what follows, [16, Sect. 3] provides an accessible exposition and further references.

## 3. Algebraic relations between values of dynamical variables under quantization

Suppose that a quantum system, hypothetically, has a sharp phase space location $(x, p)$. (We focus on a system in one spatial dimension for simplicity, the generalization to higher dimensions is straightforward.) Then all its dynamical variables $A : (x, p) \mapsto A(x, p)$, which we may assume to be the elements $A$ of some suitably chosen set of functions, e.g. the measurable functions on phase space, have sharp values $A(x, p)$. The algebraic relations between these functions, trivially, determine the algebraic relations between their values, i.e. we may assume, in parallel with KS2:

$$A(x,p) + B(x,p) = C(x,p) \text{ if and only if } v(A) + v(B) = v(C) \,, \tag{3}$$

$$A(x,p)\, B(x,p) = C(x,p) \text{ if and only if } v(A)\, v(B) = v(C) \,. \tag{4}$$

Unlike in KS2, considerations about commutativity do not arise at this stage because we have not yet considered promoting the functions $A$ to self-adjoint operators on a Hilbert space.

Quantization, as noted in the introduction, is the procedure of mapping the functions $A$ to self-adjoint operators $\hat{A}$. According to [6, Sect. 1.1], the founders of the quantum mechanical formalism envisaged a quantization procedure that (i) is linear, (ii) maps the function that is everywhere 1 on phase space to the identity operator, (iii) commutes with functional dependencies, (iv) maps position and momentum to their Schrödinger operators, and (v) maps any two dynamical variables $A$ and $B$ with Poisson bracket $\{A, B\}$ to operators $\hat{A}$ and $\hat{B}$ whose commutator obeys

$$\widehat{\{A, B\}} = \frac{1}{i\hbar}[\hat{A}, \hat{B}] \,, \tag{5}$$

where the hat symbol $\hat{}$ denotes quantization. The most famous instance of this requirement is the canonical commutation relation

$$[\hat{X}, \hat{P}] = i\hbar\, \mathbb{I} \,, \tag{6}$$

which is indeed fulfilled by the Schrödinger operators.

However, it follows from a celebrated result by Groenewold [18] that the requirements (i)–(v), taken together, are inconsistent.

Weyl quantization, which can be developed within the framework of deformation quantization, is widely regarded as the closest available approximation to the ideal expressed in conditions (i)–(v).

The fact that requirements (i)–(v) cannot be jointly fulfilled necessitates modifications to the above requirements. A flexible framework for quantization that avoids inconsistencies is deformation quantization. In this framework, a *deformed* analogue of the Poisson bracket, written as $\{\{A, B\}\}_*$, is assigned to $\frac{1}{i\hbar}[\hat{A}, \hat{B}]$. This deformed bracket differs from the ordinary Poisson bracket by terms $O(\hbar^2)$. (See below, Eqs. (14) and (29), for examples of deformed brackets.) If we denote by $\{\widehat{\{A, B\}}\}_*$ the operator to which the deformed bracket is assigned, this means:

$$\begin{aligned}[\hat{A}, \hat{B}] &= i\hbar\{\widehat{\{A, B\}}\}_* \\ &:= i\hbar\widehat{\{A, B\}} + O(\hbar^2) \,,\end{aligned}$$

where the bracket in the second line is the ordinary Poisson bracket.

As a consequence of the difference between the deformed bracket and the Poisson bracket, operator products $\hat{A} \cdot \hat{B}$ do not in general coincide precisely with $\widehat{A \cdot B}$, the operator to which products $A \cdot B$ of phase space functions $A$ and $B$ are assigned:

$$\begin{aligned} \hat{A} \cdot \hat{B} &= \widehat{A * B} \\ &:= \widehat{A \cdot B} + O(\hbar)\,. \end{aligned} \tag{7}$$

In general, the star product does not agree with the pointwise product:

$$(A * B)(x, p) \neq A(x, p)\, B(x, p)\,. \tag{8}$$

Quantization schemes need not be one-to-one and are often many-to-one. Notably, Weyl quantization and coherent state quantization assign the same operators to real-valued phase space functions that differ only on a set of measure zero. To the extent that an assignment of sharp values to all dynamical variables assigns different values to phase space functions that are mapped to the same operator by quantization, Kochen–Specker non-contextuality is not even well-defined according to our perspective because any given self-adjoint operator does not represent a single dynamical variable. In what follows, however, we will, for the sake of the argument, grant the proponent of Kochen–Specker non-contextuality that the class of dynamical variables is chosen such that a unique or privileged dynamical variable is associated with any self-adjoint operator under consideration. This dynamical variable – the one on which an operator is mapped via *de-quantization* – is called the *symbol* of the operator with respect to the chosen quantization scheme. In our proposal, the value assignment function in the sense of Kochen–Specker is the symbol evaluated at a given phase space point,

$$v_{(x_0,p_0)}(\hat{A}) := A(x_0, p_0)\,. \tag{9}$$

Under Eq. (9), the condition (i), linearity, is generally kept in candidate quantization schemes that connect $A$ with $\hat{A}$. (But see [6, Sect. 3.7] for a counterexample.) It entails that, for any three phase space functions $A$, $B$, $C$ (whether they commute or not),

$$A(x, p) + B(x, p) = C(x, p) \text{ if and only if } \hat{A} + \hat{B} = \hat{C}\,. \tag{10}$$

Plugging this into Eq. (3), we obtain the condition KS2 (a). We conclude that, as long as quantization mappings are indeed linear, KS2 (a) is plausible as an ingredient of non-contextuality.

However, as we will see in what follows, for the most promising candidate quantization schemes it does not follow from $\hat{C} = \hat{A}\hat{B}$ that $C(x, p) = A(x, p) \cdot B(x, p)$. The reason is that, due to the difference between the star product and the pointwise product Eq. (8), for $\hat{C} = \hat{A} \cdot \hat{B} = \widehat{A * B}$, the value functions give

$$v(\hat{C}) = v(\hat{A}) * v(\hat{B}) \neq v(\hat{A}) \cdot v(\hat{B})\,, \tag{11}$$

in direct contradiction with KS2 (b). This means that, in the light of quantization, KS2 (b), in contrast with KS (a), is an implausible requirement from the start.

More generally, the algebraic constraints KS2(a) and KS2(b) can be derived using the principle of "FUNC" [16]:

> Let $\hat{A}$ be a self-adjoint operator associated with observable $A$, let $f : \mathbb{R} \to \mathbb{R}$ be an arbitrary Borel function (see [17] for more details), such that $\hat{f}(\hat{A})$ is another self-adjoint operator, and let $|\phi\rangle$ be an arbitrary state; then $\hat{f}(\hat{A})$ is associated uniquely with an observable $f(A)$ such that:
>
> $$v(\hat{f}(\hat{A}))^{|\phi\rangle} = f(v(\hat{A}))^{|\phi\rangle}. \tag{12}$$

However, in the context of deformation quantization, the symbol of $\hat{f}(\hat{A})$ is $f_*(A)$, which differs from $f(A)$. Since in our approach Eq. (11) holds, this means

$$v(\hat{f}(\hat{A}))^{|\phi\rangle} = f_*(v(\hat{A}))^{|\phi\rangle} \neq f(v(\hat{A}))^{|\phi\rangle}\,,$$

contrary to the principle FUNC, Eq. (12). It is this mismatch that arguably makes KS2(b) implausible in the light of quantization. In what follows, we give concrete examples of dynamical variables that violate KS2(b) in Weyl quantization and coherent state quantization, respectively.

## 4. Weyl quantization

Weyl quantization of a dynamical variable $A(x, p)$ is defined in terms of the Fourier transform $\bar{A}(\xi, \eta)$ as

$$\hat{A} := \int\int \bar{A}(\xi, \eta) e^{2\pi i(\xi.\hat{P} + \eta.\hat{X})}\, d\xi\, d\eta\,. \tag{13}$$

The Weyl symbols follow the non-commutative Weyl star algebra,

$$\begin{aligned} (A *_W B)(x, p) &= A(x, p) \exp\left\{ \left[ \frac{i\hbar}{2} \left( \overleftarrow{\partial_q}\overrightarrow{\partial_p} - \overleftarrow{\partial_p}\overrightarrow{\partial_q} \right) \right] \right\} \\ &= A(x, p) \cdot B(x, p) + \frac{i\hbar}{2} \{A(x, p), B(x, p)\} + O(\hbar^2)\,. \end{aligned} \tag{14}$$

Thus as indicated in Eq. (8) for arbitrary star products, the Weyl star product does not simply align with the pointwise product and in general differs from it by higher order phase space derivatives acting on the functions that are being composed via the star product, which are not necessarily zero. We will now consider the plausibility of KS2 under Weyl quantization for polynomials and for the dynamical variables that are mapped onto projection operators.

### 4.1. Polynomial dynamical variables

Weyl quantization maps polynomial dynamical variables onto polynomial operators that are symmetrized in $\hat{X}$ and $\hat{P}$, i.e.

$$\widehat{x^j p^k} = \frac{1}{(j+k)!} \sum_{\sigma \in S_{j+k}} \sigma(\hat{X}, \hat{X}, \ldots, \hat{X}, \hat{P}, \hat{P}, \ldots, \hat{P}) \,, \tag{15}$$

where $S_{j+k}$ is the set of all permutations of $j+k$ objects. Alternatively, Weyl quantization for polynomial dynamical variables can be characterized uniquely by the fact that polynomials of the form $(ax+bp)^j$, where $a$ and $b$ are real numbers, are mapped onto operators $(a\hat{X}+b\hat{P})^j$.

As far as polynomial dynamical variables that depend only on $x$ or only on $p$ are concerned, the assumption KS2 (b) of the Kochen–Specker theorem is (still) plausible in Weyl quantization. To see this, consider the operators $\hat{A} = a\hat{X}^m$ and $\hat{B} = b\hat{X}^n$, with $a$ and $b$ real numbers, and $m$ and $n$ natural numbers, so that $\hat{C} = \hat{A}\,\hat{B} = ab\hat{X}^{m+n}$. The dynamical variables that are promoted to these operators in Weyl quantization (their “Weyl symbols”) are, respectively, $A(x,p) = ax^m$, $B(x,p) = bx^n$, and $C(x,p) = abx^{m+n} = A(x,p)\,B(x,p)$. Accordingly, if $A$ and $B$ are polynomials in position only (or, by analogous reasoning, in momentum only), in Weyl quantization it *does* follow from $\hat{C} = \hat{A}\hat{B}$ that $C(x,p) = A(x,p) \cdot B(x,p)$.

However, this does not apply to dynamical variables that are (non-trivially) polynomials in both position and momentum, and this already makes the assumption KS2 (b) of the Kochen–Specker theorem implausible. A simple example suffices to establish this claim, e.g. the dynamical variable

$$A(x,p) = B(x,p) = x \cdot p \,. \tag{16}$$

Weyl quantization maps this to the symmetrized operator:

$$\hat{A} = \hat{B} = \frac{1}{2}(\hat{X}\hat{P} + \hat{P}\hat{X}) \,. \tag{17}$$

The product of the operators $\hat{A} = \hat{B}$, so defined, is:

$$\begin{aligned} \hat{C} &= \hat{A}\hat{B} \\ &= \frac{1}{4}(\hat{X}\hat{P} + \hat{P}\hat{X})^2 \\ &= \hat{X}^2\hat{P}^2 - 2i\hbar\hat{X}\hat{P} - \frac{\hbar^2}{4} \,, \end{aligned} \tag{18}$$

where the third line has been written in standard order (position operator to the left of momentum operator in all product terms).

Standard order allows us to compare this operator $\hat{C}$ with the operator to which the observable $A(x,p)\,B(x,p) = (x\,p)^2$ is promoted by Weyl quantization, namely:

$$\begin{aligned} \widehat{A\,B} &= \frac{1}{6}(\hat{X}^2\hat{P}^2 + \hat{X}\hat{P}\hat{X}\hat{P} + \hat{X}\hat{P}^2\hat{X} + \hat{P}\hat{X}^2\hat{P} + \hat{P}\hat{X}\hat{P}\hat{X} + \hat{P}^2\hat{X}^2) \\ &= \hat{X}^2\hat{P}^2 - 2i\hbar\hat{X}\hat{P} - \frac{\hbar^2}{2} \,, \end{aligned} \tag{19}$$

Comparison between Eqs. (18) and (19) shows that they are not the same but differ by $-\frac{\hbar^2}{4}$. The moral is that, if $x$ and $p$ both (hypothetically) have sharp values, and if $\hat{A}$ and $\hat{B}$ represent the dynamical variable $A(x,p) = B(x,p) = x \cdot p$, then the Weyl operator $\hat{C} = \hat{A}\hat{B}$ represents the dynamical variable $(x\,p)^2 + \frac{\hbar^2}{4}$, *not* $A(x,p)\,B(x,p) = (x\,p)^2$. Accordingly, KS2 (b) is not in general plausible under Weyl quantization.

### 4.2. Projection operators

The proof of the Kochen–Specker theorem does not rely on self-adjoint operators that are polynomials of position and momentum but on projection operators. Here we show that, under Weyl quantization, KS2 (b) is not in general plausible for projection operators either.

Projection operators $\hat{\Pi}$ are characterized by the property $\hat{\Pi}^2 = \hat{\Pi}$. Phase space functions $f(x,p)$ which match this property in that $(f(x,p))^2 = f(x,p)$ are restricted to having 0 and 1 as their values. Those are the (characteristic) functions $\chi_\Delta(x,p)$ that take on the value 1 in some set $\Delta$ of phase space points and 0 elsewhere. (The mathematical features that one requires $\Delta$ to have, e.g. measurability, will depend on the set of phase space functions that one identifies with the dynamical variables.) Property KS2 will be plausible as a requirement for the values assigned to projection operators only if the dynamical variables mapped to projection operators by the quantization procedures are of the form $\chi_\Delta(x,p)$. We will see that, in Weyl quantization, this holds for some, but not all, projection operators.

To obtain the Weyl symbols of any operators, we need the inverse of Weyl quantization, the so-called Wigner transform, which is given by

$$A(x,p) = 2\int_{-\infty}^{\infty} e^{-2ipy/\hbar}\langle x+y|\hat{A}|x-y\rangle\, dy\,. \tag{20}$$

Let us first consider projection operators that project on subspaces spanned by eigenstates $|q\rangle$ of position with corresponding position eigenvalues in some range $\Delta$:

$$\hat{\Pi}_{q,\Delta} = \int_{-\infty}^{\infty} dq\ \chi_\Delta(q)\,|q\rangle\,\langle q|\,. \tag{21}$$

Since the characteristic function $\chi_\Delta(q)$ only takes two values, namely, $\chi_\Delta(q)=1$ for $q\in\Delta$ and $\chi_\Delta(q)=0$ for $q\notin\Delta$ and the different $|q\rangle\,\langle q|$ are pairwise orthogonal, this operator is indeed idempotent. The Weyl symbol of this projection operator can be calculated using the Wigner transform as follows,

$$\begin{aligned}\Pi_{q,\Delta}(x) &= 2\int_{-\infty}^{\infty} \chi_\Delta(q)e^{-2ipy/\hbar}\,\langle x+y\,|q\,\rangle\,\langle q\,|x-y\,\rangle\ dq\ dy,\\ &= 2\int_{-\infty}^{\infty} \chi_\Delta(q)e^{-2ipy/\hbar}\ \delta(x-q+y)\delta(x-q-y)\ dq\ dy,\\ &= 2\int_{-\infty}^{\infty} \chi_\Delta(q)\frac{e^{2ip(x-q)/\hbar}}{2}\delta(x-q)\ dq,\\ &= \chi_\Delta(x)\,.\end{aligned} \tag{22}$$

Above, in the second line we have used the orthogonality of the position eigenstates, and to obtain the expression in the third line we have used the relation $\int_{-\infty}^{\infty} e^{-2ipy/\hbar}\delta(\xi-y)\delta(\xi+y)dy = e^{2ip\xi/\hbar}\delta(\xi)/2$ which can be derived by writing the Dirac-delta function as a limit of a Gaussian. A similar calculation for momentum projection operator $\hat{\Pi}_{\bar{p},\Delta} = \int_{-\infty}^{\infty} d\bar{p}\ \chi_\Delta(\bar{p})\,|\bar{p}\rangle\,\langle\bar{p}|$ yields for its Weyl symbol $\Pi_{\bar{p},\Delta}(p) = \chi_\Delta(p)$.

The upshot of this calculation is that the dynamical variables mapped onto the projectors $\hat{\Pi}_{q,\Delta}$ and $\hat{\Pi}_{\bar{p},\Delta}$ are indeed of the form $\chi_\Delta$, hence it is plausible to assume, as follows from Kochen–Specker non-contextuality, that their only possible values are 0 and 1.

However, consider now a different example, namely, a canonical coherent state $|(x,p)\rangle$ of the harmonic oscillator centred around $(x,p)$ in phase space. In position space, it is given by:

$$\langle x'\mid x,p\rangle = \left(\frac{1}{\pi\, l^2}\right)^{1/4}\exp\left[-\frac{(x'-x)^2}{2l^2}+\frac{i}{\hbar}\,p\left(x'-\frac{x}{2}\right)\right]\ , \tag{23}$$

where $l=\sqrt{\hbar/(m\omega)}$, with $m$ and $\omega$ being the mass and frequency parameters in the harmonic oscillator. (Outside the context of the harmonic oscillator, one may choose an alternative length scale $l$, e.g. the Compton length in the context of atomic physics).

Its associated projection operator, written using bra-ket notation, is $|(x,p)\rangle\langle(x,p)|$. Consider further the choice

$$\hat{A} = \hat{B} = \hat{C} = |(x,p)\rangle\langle(x,p)|\,. \tag{24}$$

Trivially these three operators commute in virtue of being identical. Moreover, because of the projection property, $\hat{A}\hat{B} = \hat{A}^2 = \hat{A} = \hat{C}$. We can now consider the dynamical variable that is promoted to this operator $|(x,p)\rangle\langle(x,p)|$ in Weyl quantization — its Weyl symbol. The Wigner transform of $|(x,p)\rangle\langle(x,p)|$ is the Gaussian

$$\eta_{(x,p)}(x',p') = \frac{1}{\pi l\hbar}\exp\left\{-\frac{1}{l^2}(x'-x)^2-\frac{l^2}{\hbar^2}(p'-p)^2\right\}\ . \tag{25}$$

Evidently, as a Gaussian, $\eta_{(x,p)}(x',p')$ is not of the form $\chi_\Delta(x',p')$. Hence, $\left(\eta_{(x,p)}(x',p')\right)^2 \neq \eta_{(x,p)}(x',p')$. It follows that, while $\hat{A}\hat{B} = \hat{C}$ if $\hat{A} = \hat{B} = \hat{C} = |(x,p)\rangle\langle(x,p)|$, it is *not* true that $A(x,p)\,B(x,p) = C(x,p)$ for their Weyl symbols. Accordingly, KS2 (b) is not plausible in general for projection operators interpreted as Weyl operators. This reaffirms the conclusion already obtained above that Kochen–Specker non-contextuality, in its generality at least, is implausible under Weyl quantization.

## 5. Coherent state quantization

Coherent state quantization, which, just like Weyl quantization, can be developed in the framework of deformation quantization, may be somewhat less popular than Weyl quantization when it comes to non-relativistic quantum mechanics, but is a far more widely applicable and versatile approach to quantization [19–23]. In coherent state quantization, dynamical variables $A(x,p)$ are promoted to the so-called Toeplitz operators, which in ordinary quantum mechanics are given by

$$\hat{A} = \frac{1}{2\pi\hbar}\int A(x,p)|(x,p)\rangle\langle(x,p)|dx\,dp, \tag{26}$$

where $|(x,p)\rangle\langle(x,p)|$ are coherent state projectors. Coherent state quantization maps polynomial dynamical variables to anti-normally ordered self-adjoint operators, i.e. operators in which, when position and momentum operators are expressed in terms of creation and annihilation operators

$$\hat{a}^\dagger = \frac{1}{\sqrt{2}}\left(\frac{1}{l}\hat{X} - i\frac{l}{\hbar}\hat{P}\right) \tag{27}$$

$$\hat{a} = \frac{1}{\sqrt{2}}\left(\frac{1}{l}\hat{X} + i\frac{l}{\hbar}\hat{P}\right) , \tag{28}$$

annihilation operators appear to the left of creation operators in all product terms . Anti-normal order is also referred to as Anti-Wick order, which is why coherent state quantization as applied to ordinary quantum mechanics is also referred to as Anti-Wick quantization.

The Anti-Wick star product, which gives the Anti-Wick quantization, is given as

$$\begin{aligned} A(\alpha,\alpha^*) *_{aW} B(\alpha,\alpha^*) &= A(\alpha,\alpha^*)e^{-\overleftrightarrow{\partial_{\alpha^*}\partial_\alpha}} B(\alpha,\alpha^*) , \\ &= A(\alpha,\alpha^*)\cdot B(\alpha,\alpha^*) - \partial_{\alpha^*}A(\alpha,\alpha^*)\,\partial_\alpha B(\alpha,\alpha^*) + O(\hbar^2) . \end{aligned} \tag{29}$$

In the expression above, $\alpha = 1/\sqrt{2}\,(x/l - ilp/\hbar)$ is the complex phase space variable. Just like the case of Weyl star product, the Anti-Wick star product in general does not align with the pointwise product and thus leads to the violation of the algebraic relations underlying KS2(b), as explained above.

### 5.1. Polynomial dynamical variables

A simple example which can be used to illustrate the difference between Weyl and coherent state/Anti-Wick quantization and which also undermines the plausibility of KS2 (b) is the dynamical variable $x^2$. Using the complex variable this can be expressed as $x^2 = l^2/2\,(\alpha+\alpha^*)^2$. This allows straightforward mapping to the anti-normally ordered operator

$$\begin{aligned} \widehat{(x^2)} &= \frac{l^2}{2}\left(\hat{a}^2 + 2\hat{a}\hat{a}^\dagger + (\hat{a}^\dagger)^2\right) \\ &= \frac{l^2}{2}\left(\hat{a}^2 + \hat{a}\hat{a}^\dagger + \hat{a}^\dagger\hat{a} + (\hat{a}^\dagger)^2 + 1\right) \\ &= \hat{X}^2 + \frac{l^2}{2} , \end{aligned} \tag{30}$$

where creation and annihilation operators have been eliminated in the third line to facilitate comparison with the result from Weyl quantization, namely, $\hat{X}^2$. This example already shows that the operators $\hat{A} = \hat{B} = \hat{X}$ and $\hat{C} = \hat{X}^2$ form a counterexample to the plausibility of KS2 (b) in coherent state quantization. These operators commute with each other, they fulfil $\hat{A}\hat{B} = \hat{C}$, yet, for their Anti-Wick symbols $A(x,p)\,B(x,p) = x^2 = C(x,p) - \frac{l^2}{2} \neq C(x,p)$ (since $l \neq 0$). An analogous result is obtained for $\hat{A} = \hat{B} = \hat{P}$ and $\hat{C} = \hat{P}^2$. Unlike in Weyl quantization, we do not have to consider polynomials that include non-trivial products of both $x$ and $p$ to discover that KS2 (b) is implausible in coherent state quantization.

### 5.2. Projection operators

We now turn to projection operators in the image of coherent state quantization. For any self-adjoint operator that has both a Weyl and an Anti-Wick symbol, the former is obtained from the latter by a Weierstrass transformation, which is a "smoothing" in form of a convolution with the Gaussian $\eta_{0,0}$ as introduced in Eq. (25) [20, p. 141]. It follows that the projection operators $\hat{\Pi}_\Delta$, where $\Delta$ is some range of positions or momenta, are not in the image of coherent state quantization, i.e. they are not Anti-Wick operators: As seen above, the Weyl symbols of the projections $\hat{\Pi}_\Delta$ onto subspaces spanned by position or momentum eigenstates are of the form $\chi_\Delta$, and these functions, which are discontinuous to the extent that they are non-trivial, are plainly not obtained as "smoothings" from any phase space function $A(x,p)$ via convolution with a Gaussian.

This gives an interesting twist to the question of whether Kochen–Specker non-contextuality is plausible under coherent state quantization, namely, that, in coherent state quantization, we cannot in general assume for self-adjoint operators that they represent dynamical variables in the first place. Notably, not all projection operators represent dynamical variables. The proof of the Kochen–Specker theorem, which demonstrates that it is impossible to assign sharp values to arbitrary projection operators, therefore, does not speak to the viability of assigning sharp values to all dynamical variables under coherent state quantization.

There is a parallel here between the present considerations and those underlying the "nullification" of the Kochen–Specker theorem claimed by Meyer, Clifton, and Kent due to the inevitable finite precision of real-life measurements [24,25]. In the models proposed by those authors, not all self-adjoint operators are assigned sharp values, and for those that are assigned sharp values, Kochen–Specker non-contextuality is fulfilled. However, the reasoning that Meyer, Clifton, and Kent give to justify withholding sharp values from certain self-adjoint operators is very different from the reasoning provided here. Those authors do not question that all self-adjoint operators qualify as dynamical variables. But they argue that, in view of the finite precision of actual measurements, it suffices to assign sharp values to only some of them (Appleby [26] calls this doctrine "existential contextuality"), provided that these form a dense subset of the overall set of self-adjoint operators. Here, in contrast it is argued that some self-adjoint operators, notably, the projectors on subspaces spanned by position and momentum eigenstates, just do not represent any dynamical variables under coherent state quantization, simply because they are not in the image of coherent state quantization.

It is worthwhile to note, however, that some projection operators do have Anti-Wick symbols, i.e. they do represent dynamical variables under coherent state quantization (at least, if one accepts the Dirac delta-distribution as a dynamical variable), namely, the projectors $|(x,p)\rangle\langle(x,p)|$ onto coherent states. A look at the coherent state quantization map Eq. (26) shows that one can take as the Anti-Wick symbol of the coherent state projector the normalized delta-distribution $2\pi\hbar\,\delta\left(x'-x, p'-p\right)$ (provided one accepts it as a dynamical variable despite not being a proper function). Weyl quantization, in contrast, does not apply to this delta-distribution $2\pi\hbar\,\delta(x'-x,p'-p)$.[3] This indicates an important difference between Weyl quantization and coherent state quantization: In Weyl quantization, the yes/no-question "Is the system at the exact point $(x,p)$ in phase space?" is "forbidden" in the sense that the dynamical variable corresponding to it is not represented by any self-adjoint operator, whereas it is "allowed" in coherent state quantization! However, since $2\pi\hbar\,\delta(x'-x,p'-p)$ does not have the form of a characteristic function, we see that consideration of projectors $|(x,p)\rangle\langle(x,p)|$ onto coherent states in coherent state quantization does not provide any support for the plausibility of KS2 (b) either.

In the next section we will see that it is very natural to combine coherent state quantization with the idea that all dynamical variables have sharp values at any given instant.

## 6. Coherent state quantization and the Husimi function

In this section, we point out that combining coherent state quantization with assigning sharp values to all dynamical variables leads to an interpretation of the Husimi Q-function as a proper probability density on phase space, as recently proposed based on different considerations by Drummond and Reid [9] and Friederich [10]. This can be seen as follows:

In the phase space formulation of quantum mechanics, the quantum expectation value $Tr(\hat{A}\hat{\rho})$ associated with a self-adjoint operator $\hat{A}$ can be computed in a variety of different ways as a weighted phase space integral [27, Eq. (2.1)]:

$$\begin{aligned}\langle\hat{A}\rangle &= Tr\left(\hat{A}\hat{\rho}\right)\\ &= \int A_O(x,p)\,F_O(x,p)\,dx\,dp\,.\end{aligned} \tag{31}$$

The index $O$ stands for different types of operator ordering or, alternatively, different symbols, associated with different quantization procedures. The most often used symbols are the Weyl symbol (corresponding to symmetric operator ordering for polynomials), the Wick symbol (normal ordering), and the Anti-Wick symbol (anti-normal ordering). The quantity $F_O(x,p)$ functions similarly to a probability distribution — notably its phase space integral is normalized to 1 — with the important caveat that it can be negative in some regions of phase space (whence it is sometimes referred to as a "quasi-probability distribution".) The choice of function $F_O$ must be made such that it matches the chosen symbol type. Notably, the Wigner function must be selected to match the Weyl symbol, the Glauber–Sudarshan P-function must be selected to match the Wick symbol, and the Husimi Q-function must be selected to match the Anti-Wick symbol. The Husimi function is defined as

$$Q(x,p) = \frac{1}{\pi}\langle(x,p)|\hat{\rho}|(x,p)\rangle. \tag{32}$$

Unlike the Wigner and Glauber–Sudarshan functions, it is strictly non-negative and, as such, has all the formal features of a probability distribution.

Clearly, if the dynamical variable $A$ is promoted to some self-adjoint operator $\hat{A}$ by coherent state quantization, then $A$ is $\hat{A}$'s Anti-Wick symbol. In the light of Eq. (31) this means that, under coherent state quantization, any quantum expectation value $Tr\left(\hat{A}\hat{\rho}\right)$ can be written as "classical" phase space integral, where the Husimi function $Q(x,p) = F_{\text{Anti-Wick}}(x,p)$ associated with $\hat{\rho}$ plays the role of the probability density, that is:

$$Tr\left(\hat{A}\hat{\rho}\right) = \int A(x,p)\,Q(x,p)\,dx\,dp\,, \tag{33}$$

where $A(x,p)$ is the actual dynamical variable represented by $\hat{A}$, not just some auxiliary quantity.

This observation suggests that embracing coherent state quantization in combination with interpreting the Husimi function as a genuine phase space probability distribution may be a promising avenue towards ascribing sharp values to all dynamical variables in an organic way. As observed by Friederich [10, Sect. 5], this approach does not fit within the ontological models framework. That framework, however, is used by Spekkens [28] to derive a no-go theorem which rules out hidden variable models with positive semi-definite quasi-probability distributions over phase space. That theorem, therefore, does not speak to the viability of interpreting the Husimi function as a phase space probability distribution in combination with coherent state quantization.

Nevertheless, one may wonder whether an approach centred around the Husimi function rather than the Wigner function can possibly be empirically adequate. [29] judge that the overall body of empirical results in atomic and molecular physics is consistent with Hamiltonians obtained via coherent state quantization just as much as Weyl quantization, at least in the non-relativistic regime. [9,10] argue that, if interpreting the Q-function as a proper probability distribution is to yield an empirically adequate account, measurement outcomes cannot possibly reflect pre-existing values of variables. Rather, they must be the products of non-trivial interactions between measured systems and measuring instruments.

Bell once summed up similar insights with respect to Bohmian mechanics:

[3] To see this, recall that, for any operator that has both a Weyl and an Anti-Wick symbol, the former is obtained from the latter via a Weierstrass transformation. But $2\pi\hbar\,\delta(x'-x,p'-p)$ is not obtained from any phase space function via a Weierstrass transformation. Since we can construct Anti-Wick operators for a very wide class of phase space functions via Eq. (26), we can conclude that $2\pi\hbar\,\delta(x'-x,p'-p)$ is not the Weyl symbol of any self-adjoint operator.

> The result of a 'spin measurement', for example, depends in a very complicated way on the initial position $\lambda$ of the particle and on the strength and geometry of the magnetic field. Thus the result of the measurement does not actually tell us about some property previously possessed by the system, but about something which has come into being in the combination of system and apparatus. [30, p. 35]

However, unlike Bohmian mechanics, the account that combines coherent state quantization with a sharp assignment of values to all dynamical variables allows one to preserve a close conceptual link between dynamical variables and the self-adjoint operators assigned to them via quantization. To see how Bohmian mechanics severs that link, recall that, in Bohmian mechanics (see [31] for an introduction), the phase space probability distribution $F(x,p)$ follows from the assumption that $|\psi(x)|^2$ is the probability distribution in position space together with the fact that momenta are fixed according to the guidance equation

$$p(x,t) = m\partial_t x(t) = \hbar \,\mathrm{Im}\left(\frac{\nabla\psi}{\psi}\right)(x,t)\,. \tag{34}$$

Clearly, for wave functions $\psi(x)$ whose imaginary parts vanish, e.g. the ground states of the harmonic oscillator and the hydrogen atom, the Bohmian momentum $p(x,t)$ is zero. Accordingly, the expectation values of all multiples $p^n$ of momentum, including the expectation value of $p^2$, are also zero in these states. At the same time, the expectation value of the squared Weyl operator of momentum $\hat{P}^2$, calculated in the standard quantum mechanical way, is nonzero in these states. Accordingly,

$$\begin{aligned}\langle p^2\rangle &= \int p^2\, F(x,p)\, dx\, dp \\ &\neq Tr(\hat{P}^2\hat{\rho})\,.\end{aligned} \tag{35}$$

Thus, in general, in Bohmian mechanics

$$\langle A\rangle = Tr(\hat{A}\hat{\rho})\,, \tag{36}$$

does *not* hold, whether one takes $\hat{A}$ to be the operator to which $A$ is promoted by Weyl quantization, coherent state quantization, or some other quantization scheme. In contrast, as we have seen, the "Anti-Wick version" of Eq. (36) *does* hold in the account that combines coherent state quantization with assigning sharp values to all dynamical variables.

## 7. Conclusion

In this paper we have argued that, for quantum theories obtained by quantizing some classical theory, Kochen–Specker non-contextuality was never a reasonable assumption to begin with when attempting to assign sharp values to all dynamical variables in quantum mechanics. This would only have been the case if algebraic relations between dynamical variables had been invariant under quantization. However, as we have shown using various examples of polynomial dynamical variables and projection operators in Weyl and coherent state quantization, this is not in general the case. As a general lesson from our considerations, we suggest that *dequantization* may be a generally attractive route for assigning sharp values to self-adjoint operators. Such assignments, if they are to be empirically adequate, will inevitably violate Kochen–Specker non-contextuality, but they can be mathematically natural, as the Weyl and coherent state (de-) quantizations arguably are.

## CRediT authorship contribution statement

**Simon Friederich:** Writing – review & editing, Writing – original draft, Supervision, Methodology, Conceptualization. **Mritunjay Tyagi:** Writing – review & editing, Writing – original draft.

## Declaration of competing interest

The authors declare the following financial interests/personal relationships which may be considered as potential competing interests: Simon Friederich reports financial support was provided by Netherlands Organization for Scientific Research. If there are other authors, they declare that they have no known competing financial interests or personal relationships that could have appeared to influence the work reported in this paper.

## Acknowledgements

We would like to thank Emily Adlam, Sean Gryb, Karim Thebault as well as audience members of the 4th Dutch History and Philosophy of Physics workshop, Nijmegen 2024, and Foundations 2025, Gdansk, for useful discussions.

This research was funded by the Netherlands Organization for Scientific Research (NWO), project VI.Vidi.211.088.

## Data availability

No data was used for the research described in the article.